\documentstyle[12pt]{article}
\input{epsf.sty}
\def\plotone#1{\centering \leavevmode
\epsfxsize=.95\columnwidth \epsfbox{#1}}

\newcommand{\micron}{$\mu$m}
\newcommand{\lsun}{L{\small $\odot$} }

\begin{document}

\title{Infrared emission from interstellar dust cloud with two 
embedded sources: IRAS 19181+1349 }

\author{A.D. Karnik and  S.K. Ghosh}
\maketitle
\begin{center}
Tata Institute of Fundamental Research \\
 Homi Bhabha Road, Colaba \\
Mumbai (Bombay) 400 005, India
\end{center}

\newpage
\begin{abstract}
 Mid and far infrared maps of many Galactic star forming regions
show multiple peaks in close proximity , implying more than one
embedded energy sources. With the aim of understanding
such interstellar clouds better, the present study models the
case of two embedded  sources. A radiative transfer scheme
has been developed to 
deal with an uniform density dust cloud in a cylindrical geometry,
which includes
isotropic scattering  in addition to the 
emission and absorption processes. 
This scheme has been  
applied  to the Galactic star forming region
associated with IRAS 19181+1349,
which shows observational evidence for two embedded energy sources.
Two independent modelling approaches have been adopted, viz., to fit
the observed spectral energy distribution (SED) best; or 
to fit the various radial profiles best, as a function of wavelength.
Both the models imply remarkably similar physical parameters.
\end{abstract}
\newpage

\section{Introduction}
 Galactic star forming regions mostly comprise of
Young Stellar Objects (YSOs) / protostars still
buried inside / in the vicinity of the parent
interstellar cloud from which they are formed.
Hence the study of YSOs leads to understanding of
the interstellar medium in the close neighborhood
of the starbirth. Early evolution of star forming 
regions is obviously more important from the point of 
understanding the star formation process itself. 
Typically the photons from the embedded energy source,
protostar / ZAMS star, get reprocessed by the
dust component of the interstellar medium in the immediate
neighborhood. The dust grains absorb / scatter the
incident radiation, depending on their dielectric
properties. The dust grains acquire an equilibrium 
temperature based on the local radiation field which
depends on the distance from the energy source and
secondary radiation from other grains. The reradiation
from the grains in the outer regions, is what is
observable. Hence, the emerging observable spectrum
can in principle be connected  to the spectrum 
of the embedded protostar / ZAMS star through detailed
radiation transfer provided some details about the
geometry are known.

In order to study the earliest
 stage of star formation, a large amount of observational 
 effort is directed towards far infrared / sub-mm 
 observations of prospective young star forming regions.
 Many Galactic star forming regions have been mapped
 with near diffraction limited angular resolutions, using
 Kuiper Airborne Observatory, Infrared Space Observatory (ISO),
 James Clerk Maxwell Telescope etc. in recent times.
 In many cases, the maps of continuum emission
 resolve several closeby individual intensity peaks
 as evident from the morphology of their isophots.
 
  The present study is a step in the direction of
  extracting maximum possible information about the
  geometrical and physical details of the source
  by comparing radiative transfer models with observations
  in  cases where two nearby sources are resolved.
  Here, the ``nearby" implies the interference in 
  energetics of each of the two resolved source by
  the other one. 
The heating of dust in a cloud with multiple sources
has been studied by Rouan (1979) under certain simplification.
 However, the problem of more than one embedded source
has rarely been addressed  quantitatively.
The existing observational data showing clear evidence
of resolved multiple embedded sources justify the need to
explore geometries dealing with more than one embedded source.
Ghosh \& Tandon (1985)
attempted to study the case
of two embedded sources with many simplifying assumptions.
They neglected some basic phenomena like scattering which
limited its applicability to $\lambda \geq 50$ \micron\
 only.
The present work is an extension of this earlier 
attempt by including the effect of isotropic scattering.

 In section 2 the problem has been formulated and the
radiation transfer scheme is described. 
In section 3, we model observations of IRAS 19181+1349
which shows evidence of having two embedded sources.
The results of our modelling are then discussed.

\section{Model Formulation}

 The primary aim of the present  model is to reproduce the
infrared emission from a star forming cloud with two embedded sources,
keeping the computational complexities at a minimum level.

The interstellar cloud is assumed to be of cylindrical shape.
As a starting point, an uniform density of the cloud has been
assumed. The line joining the two embedded ZAMS stellar / 
protostellar energy sources defines the symmetry axis of the
problem. 
Around each of the sources there will be  a dust free cavity  
(Fig. 1).
The existence of such a cavity is widely accepted due to evaporation of
the dust grains in the intense radiation field. In addition, radiation
pressure effects on the dust grains may also play a role
in deciding the cavity size.
The radiation transfer is carried out through the dust component alone. 
Dust grains with a continuous size distribution have been considered, 
and their
composition is a parameter of modelling. Three types of grains, viz.,
Graphite, Astronomical Silicate and Silicon Carbide have been invoked
since their existence is generally accepted.
All properties of the dust grains, viz., absorption and scattering
coefficients as a function of wavelength, for various sizes and 
all three types of grains,
have been taken from Laor \& Draine (1993).
The size distribution of  all the three types
of grains has been  assumed to be power law ($n(a)da\sim a^{-3.5}da$)
as per Mathis, Rumpl \& Nordsieck (1977).
The wavelength grid used 89 points covering
from the Lyman continuum limit to the millimeter wavelengths.

The geometrical parameters
relevant to the   model
(Fig. 1) include the radius of the cylinder (R$_{cyl}$),
radii of dust free cavities near the two sources ( R$_{c1}$, R$_{c2}$),
 and distance
between two exciting sources (D). Other physical parameters are
the  composition of the dust (relative abundances of three 
types of dust components)
and the dust density expressed in optical depth at 100 \micron\ ($\tau_{100}$).
The optical depth at any other wavelength
is uniquely connected to $\tau_{100}$ via the dust
properties and composition assumed in the perticular model.
The cylindrical interstellar cloud is divided into $n_z$ identical
discs. Each of these discs are further divided into $n_r$
annular rings.
The $n_z$ is chosen such that each disc is optically
thin even at UV,  along the z-axis.
Along the radial direction (total number of
grid points being $n_r$), a two stage
grid has been employed which is initially
nearly logarithmically spaced (near the symmetry axis) and linearly spaced
in the outer regions of the cylinder. This scheme of radial grid
has been arrived at by keeping the optical depth related inaccuracies 
under check, for the entire wavelength region considered for the 
radiation transfer.
Both the near logarithmic and the linear grids are matched 
by ensuring that radial cell size, $\delta r(n_r)$, is a smooth 
function of $n_r$.
For modelling attempts of IRAS 19181+1349 a  grid of 600 points in axial
direction and of  25 points in radial direction were employed.

The relevant calculations  are 
represented in equations \ref{equgeom}-\ref{equpowabs}. 
For clarity the frequency suffix has been
dropped from all the terms, though calculations are performed for
each frequency grid point. The quantities in angled brackets
are averages over the dust size distribution.
The code is simplified and optimized in many ways in view
of the memory requirements and speed. 
Initially, factors
totally dependent on the  geometry of the problem and which
do not change in each iteration are calculated. These
include the optical depth terms and the geometric integrals involved in
computations of  radiation received by an unit volume
element of the ring $i$ from  the ring $j$ (equation \ref{equgeom}).
 The geometric
symmetry is such that such terms only depend on the axial separation
between the two rings and their respective radii. The total flux
absorbed and scattered by unit volume in each ring due to
radiation from other rings is then calculated (equation \ref{equring}).
Also, the radiation received by
an unit volume element of the ring $i$ from embedded sources depends
only on the geometry and the optical depth per unit length of the
particular model and hence is fixed over iterations (equation \ref{equsrc}). 
The equilibrium temperature of the dust grains (on the median 
circle) for any particular annular ring is calculated 
using an iterative scheme by 
equating the power radiated by the dust (equation \ref{equpowemit}) to the
power absorbed (equation \ref{equpowabs}). 
The latter is contributed by the embedded exciting sources 
(attenuated by the line of sight dust)
as well as secondary emission and scattering from dust grains
in all other annular rings (equation \ref{equflux}).
This simplifies the calculations leading to a set of coupled equations
with only two parameters per ring, temperature ( $T_i$ ) and   $F^{i}$ changing
from iteration to iteration thus greatly reducing the memory requirements .

\begin{equation}
F_{r}^{i,j} =
\sum_{\theta}\frac{ e^{-<\pi a^2 Q_{ext}>n_d d_{i,j,\theta} }}{d_{i,j,\theta}^2} 
[ <\pi a^2 Q_{abs}> n_d B( T_j) + 
\frac {< \pi a^2 Q_{scat} > F^{j} }{ 4 \pi < \pi a^2 Q_{ext}> } ]
\Delta r_j \Delta z \Delta \theta r_j
\label{equgeom}
\end{equation}

\begin{equation}
F_{r}^{i} = \sum_{j=1,j\neq i}^{n_r \times n_z}<\pi a^2 Q_{ext}> n_d  F_{r}^{i,j}
\label{equring}
\end{equation}

\begin{equation}
F_{s}^{i} = \sum_{s=1}^2 <\pi a^2 Q_{ext}>  \frac{\pi R_s^2 F_{s}}{r_{i,s}^2 }
     e^{ -<\pi a^2 Q_{ext}>n_d ( r_{i,s} - r_{c,s}) }
\label{equsrc}
\end{equation}

\begin{equation}
F^{i} = F_{r}^{i} + F_{s}^{i}
\label{equflux}
\end{equation}

\begin{equation}
F_{e}^{i} =  <\pi a^2 Q_{abs}> 4 \pi B(T_i) 
\label{equfluxemit}
\end{equation}

\begin{equation}
P_{Emitted} =  \int F_{e}^{i} d\nu
\label{equpowemit}
\end{equation}

\begin{equation}
P_{Absorbed} =  \int \frac{<\pi a^2 Q_{abs} >}{<\pi a^2 Q_{ext}>}(F_{r}^i + F_{s}^i) d\nu
\label{equpowabs}
\end{equation}

 where \\
$d^2$  $\equiv (z_i - z_j )^2 + (r_i)^2 + (r_j) ^2 -2 r_i r_j \cos(\theta)$ \\
  $\theta$ is azimuthal angular difference between two volume units under
 consideration.\\
$r_{i,s}^2$ $\equiv (z_i-z_s)^2 + r_i^2$ \\
$r_{c,s}^2$ $\equiv $ dust cavity radius for source $s$. \\
$Q_{ext}$ $\equiv Q_{abs} + Q_{scat}$ \\
$n_d$ $\equiv$ number density of dust grains. \\
$T_{i}$ $\equiv$ Temperature of dust in ring $i$. \\
$F_{s}$ $\equiv$ surface flux spectrum for source $s$. \\
$F_{r}^{i}$ $\equiv$ Flux absorbed and scattered by unit volume of ring $i$, 
		due to other rings. \\
$F_{s}^{i}$ $\equiv$ Flux absorbed and scattered by unit volume of ring $i$, 
		due to  sources. \\
$F^{i}$ $\equiv$ Total flux absorbed and scattered by unit volume of ring $i$. \\
$B(T)$ $\equiv$ Planck function. \\

The dust temperature for each annular ring, leading to the
temperature distribution throughout the cloud, is determined
iteratively. Initially (the very first iteration), only the 
two embedded sources power the heating of the grains. From the
second iteration onwards the effects of secondary heating and
scattering are taken into account.
In this iterative procedure the temperature of each annular ring is
gradually updated in each iteration satisfying the  condition
 $P_{Emitted} = P_{Absorbed}$. 
 The iterations are continued till the  fractional changes in  absorbed power for 
each annular ring, between successive iterations,
 reduces below the convergence criteria.

 The emergent intensity distribution, as seen by a distant
observer is predicted by integrating the emitted and
scattered radiation along relevant lines of sight and taking
account of extinction due to the line of sight optical depth.
This  spatial intensity distribution at any selected wavelengths
is convolved with the relevant instrumental beam profile (PSF) for 
direct comparison with
observations.

 Before applying the  scheme developed above
 to any astrophysical source it is
necessary to verify its reliability and  quantify
its accuracy. 
For this,
we  simulate the case of a single embedded source by ``dimming''
one of the two sources  thereby keeping our original code intact during
the test runs.
This  simulates a single exciting source
embedded on the symmetry axis of the uniform density cylindrically shaped
cloud.
The size of the cylindrical cloud has been chosen such that the
results from other codes using spherically symmetric geometry could be
compared effectively.
We have used the well established code
 CSDUST3 (Egan, Leung \& Spagna 1988) for such a comparison.

The radius of the cylindrical cloud used for comparing  our code with the
spherically symmetric code was identical
to that of the ``equivalent" spherical cloud. The length of the 
cylinder (along z-axis) is twice this radius. 
Figure 2 is a schematic of the
cylindrical  as well as the spherical  models.

The assumed radius of the cylinder, as well as the sphere, was 2 pc.
The radius of the dust free cavity near the source was adopted to be .01 pc.
In order to make the comparison possible and effective, all
model parameters were made identical for both the codes.
These include : dust size distribution; dust number 
density (and hence total optical depth along the line of sight
between the embedded source and the distant observer). 
A typical dust composition has been assumed consisting of 
50 \% Graphite and 50 \% Astronomical Silicate.
The embedded energy source was assumed to be a single ZAMS
O5 star with luminosity $9.0\times 10^5 L_{\odot}$.
Some parameters were explored (e.g. optical depth) to study
regions of validity with specified accuracy,
by changing them identically for both the schemes.
A large range in the optical depth  $\tau_{100}$,
viz., 1.0 x $10^{-3}$ to 0.1 was covered during the
test runs.

The comparison of the emergent spectral energy distributions (SEDs) 
from both the schemes
for various optical depths are shown in Fig. 3. 
It may be  seen that the  SEDs
match quantitatively  over a wide range of optical depths
from mid infrared to millimeter
wavelengths. There are some differences at near infrared wavelengths; 
this is to be expected from the differences in the cell sizes and
the geometry. However, since the main motive of the present
study is to interpret measurements in the wavebands beyond
the mid--IR, our code  may be considered to be  satisfactory.
From this comparative study we conclude that our code
is accurate upto an optical depth corresponding to
$\tau_{100} \approx 0.06$ ($\tau_{1~\mu m} \approx 6$), 
for the present choice of grid points. 
This limit already corresponds to 
much denser clouds than generally found in the Galactic 
star forming regions.

\section{IRAS 19181+1349}
The Galactic star forming region IRAS 19181+1349 is a IRAS
Point Source Catalog (IRAS PSC) source associated with the
radio source G48.60+00. The presence of a radio continuum suggests
ongoing high mass star formation in the region.
This source has been resolved in two components
in the  210 \micron\  map (Fig. 4b) generated 
from the observations using the 
TIFR 1--meter balloon borne telescope (Karnik et al, 1999). 
Although a single IRAS PSC source is associated with this star forming
region, the HIRES processing of the IRAS survey data has led to the
resolution of these two sources in the 12 \& 25 \micron\ band maps.
The map at 12 \micron\ is shown in Fig. 4a.
We denote FIR peak with higher R.A. as S1 and that of lower R.A. S2,
respectively, hereafter.
Zoonematkermani et al. (1990) have reported four compact radio sources
48.603+0.026,48.609+0.027,48.606+0.023 and 48.592+0.044 in this region.
First three of these closely match  S1 (positionally)
while the fourth one matches S2.
Kurtz, Churchwell \& Wood (1994) have also reported three ultracompact
HII regions  which are positionally close to S1, indicating  ongoing
high mass star formation in a very  early evolutionary phase. 
This source has also been observed
using ISOCAM  
instrument onboard Infrared Space Observatory (ISO) 
using seven filters centred at four PAH features and 3 continuua
(3.30, 3.72, 6.00, 6.75, 7.75, 9.63, 11.37 \micron;
Verma et al 1999).
The ISOCAM images at all the bands show
two extended prominent complexes with multiple peaks.
Hence from radio as well as mid to far infrared observations the nature
of IRAS 19181+1349 is established to be of double embedded source/cluster type.

This seems to be an ideal astrophysical example of interstellar cloud
with two embedded sources.
We attempt to model this source with our
scheme to extract important physical parameters about 
the embedded energy sources as well as the intervening interstellar
medium in IRAS 19181+1349. Next we describe the results obtained
from such a study.

 The source IRAS 19181+1349 was considered as  a cylindrical dust
cloud with two protostellar /ZAMS stellar sources embedded along 
the axis of the cylinder. 
The size of the cloud and the dust density were used as free
parameters. The sum of the luminosities of the two 
embedded sources is determined by integrating the
observed spectral energy distribution (SED). The observations used for
this integration includes the four IRAS bands and the two TIFR bands.
This total luminosity is treated as an observational constraint
in the modelling.
Further observational constraints include :
the shape of the SED which was obtained
from HIRES--IRAS,  TIFR  and ISO observations, and the structural morphology 
of IRAS 19181+1349
as reflected by the isophote contours of the high angular resolution
maps at 12 \& 210 \micron. 
As our code does not deal with the physics of PAH emission at present, 
only the data in ISOCAM filters sampling the continuua have been
used to constrain the model.

  Two completely independent approaches have been followed in modelling
IRAS 19181+1349 using our scheme. 
In each approach, all the model parameters are floated to obtain the
best fit model.
The parameters fine tuned to achieve the best
fit model are :  luminosities
of individual embedded sources; geometrical size details of the cylindrical
cloud (including the size of the cavity); 
and the dust density / optical depth.
The main aim of the first
approach is to optimize the fit to the observed SED (hereafter $M_{SED}$). 
The second approach optimizes the one dimensional radial intensity
profiles at selected
wavelengths covering mid to far infrared bands (hereafter $M_{RC}$).
The radial profiles
are taken along geometrically interesting axes, viz., along the
line joining the two embedded sources and along lines perpendicular
to the cylinder axis and passing through either of the two sources 
(see Fig. 4).
Whereas the first approach gives precedence to the overall energetics
the latter gives more importance to the structural details in the
isophotes, particularly close to the embedded sources and the region
between them. The actual reality may lie somewhere in between these
two approaches.

The axes where
radial cuts have been taken are also displayed on the 12 \& 210 \micron\
maps in Fig. 4. 
The fits to the observed SED for both $M_{SED}$ as well as 
$M_{RC}$ are presented in Fig. 5. 
Comparison of the radial cuts at 12 \& 210 $\mu$m along
the three axes between the observed maps and the 
best fit $M_{RC}$ model are shown in Fig. 6.

The $M_{SED}$ approach fits the observed SED very well right 
through near--IR to sub-mm, though there is some discrepancy
in the far--IR (Fig. 5). On the other hand, fit from
the $M_{RC}$ approach is resonable for $\lambda \geq$ 25 \micron\ only.
The radial cuts at 210 \micron\  fit the observed
data very well. However, at 12 \micron\ although the model
predictions qualitatively agree with the data, there are
discrepancies viz., the extent of emission along vertical
axis and the relative contrast of the minima between the
two sources. From the above it appears that the $M_{RC}$
approach is very sensitive to the exact distribution of sources,
specially   at shorter wavelengths (which traces much hotter
dust and hence physically closer to the exciting source). One
possibility for these discrepancies is that our model
is too simplified than the actual reality in IRAS 19181+1349.
For example there may be a distribution of sources along the
line joining S1 and S2 and away from it, but with more
concentration near S1 and S2. This way the discrepancy for $M_{RC}$
in the mid--IR part of SED will also be explained.
Assuming a constant dust density in the cloud may also
be a major reason for the discrepancy.

 The results of modelling IRAS 19181+1349 in the form of best fit model 
 parameters as found under both the approaches ($M_{SED}$ and $M_{RC}$)
are presented in Table 1. 
The value of $\tau_{100}$ found from both the approaches
are well within the range of validity of the code,
i.e. much less than 0.06. The luminosities of both the sources are similar
and the dust is found to be predominantly (80\%) Silicate with
no Silicon Carbide, the rest (20\%) being Graphite.
It is remarkable that almost all the 
important parameters are identical as found from both the approaches.
The only difference is seen at the physical sizes of the cavities.
The remarkable similar physical parameters obtained from
both the approaches 
gives good confidence on the derived astrophysical parameters
for the Galactic star forming region IRAS 19181+1349.

This scheme will be useful
to model other similar double sources. 
A huge database from ISO (SEDs as well
as images leading to radial cuts, covering
2.5 to 200 $\mu m$ of the spectrum) will become
available soon for such modelling.

\section {Summary}

High angular resolution mid / far infrared maps of many 
Galactic star forming 
regions show evidence for multiple embedded energy sources in the
corresponding interstellar clouds. 
 With the aim of studying such star forming regions with two
embedded sources, a  scheme has been developed to carry
 out radiative transfer calculations
in a uniform density dust  cloud
of cylindrical geometry, and which includes
the effect of isotropic scattering in addition to the absorption
and emission processes. 
In addition to the luminosities of the two embedded energy sources,
the cylindrical cloud size, separation between the two sources,
dust density and composition are the 
parameters for the modelling. The accuracy of our scheme has been tested 
by comparing the results with a well established 1-D code.

An attempt was made to model the Galactic star forming region
associated with IRAS 19181+1349 which shows double peaks
in the mid and  far infrared maps using the scheme described here.
Two independent approaches were employed to find the
best fit models for IRAS 19181+1349. Whereas the first ($M_{SED}$)
approach aims to fit the observed SED best; the latter ($M_{RC}$)
aims to fit the radial intensity profiles (along a few
important axes) at mid to far infrared wavebands.
Interestingly  most of the crucial model parameters like
luminosities, effective temperatures, dust composition,
optical depth etc turn out to be identical under both the 
approaches. 

\noindent{\bf Acknowledgments}\\
It is a pleasure to thank the members of the Infrared Astronomy
Group of T.I.F.R. for their encouragement.


\begin{table}[ph]
\caption{Best fit  model Parameters for M$_{RC}$ and M$_{SED}$ schemes. 
\label{modeltab} }
\begin{center}
 \begin{tabular}{lll}
 \hline 
 Parameter  & M$_{RC}$  & M$_{SED}$  \\
 \hline 
 $R_{c}$ (pc)(Source1)         & 0.11 & 0.03 \\
 $R_{c}$ (pc)(Source2)         & 0.09 & 0.01 \\
 $L_{source1}$ (\lsun) & 6.2 $\times 10^5$  & 6.2 $\times 10^5$ \\
 $L_{source2}$ (\lsun)	& 6.3 $\times 10^5$ & 6.3 $\times 10^5$ \\
 $\tau_{100 cyl}$ (/pc) &  0.024 & 0.028  \\
 $R_{cyl}$  (pc)         &  2.1  & 1.8 \\
 Graphite:Silicate:SiC & 20:80:0 & 20:80:0 \\
 \hline
 \end{tabular}
\end{center}
 \end{table}

\begin{figure}[p]
\plotone{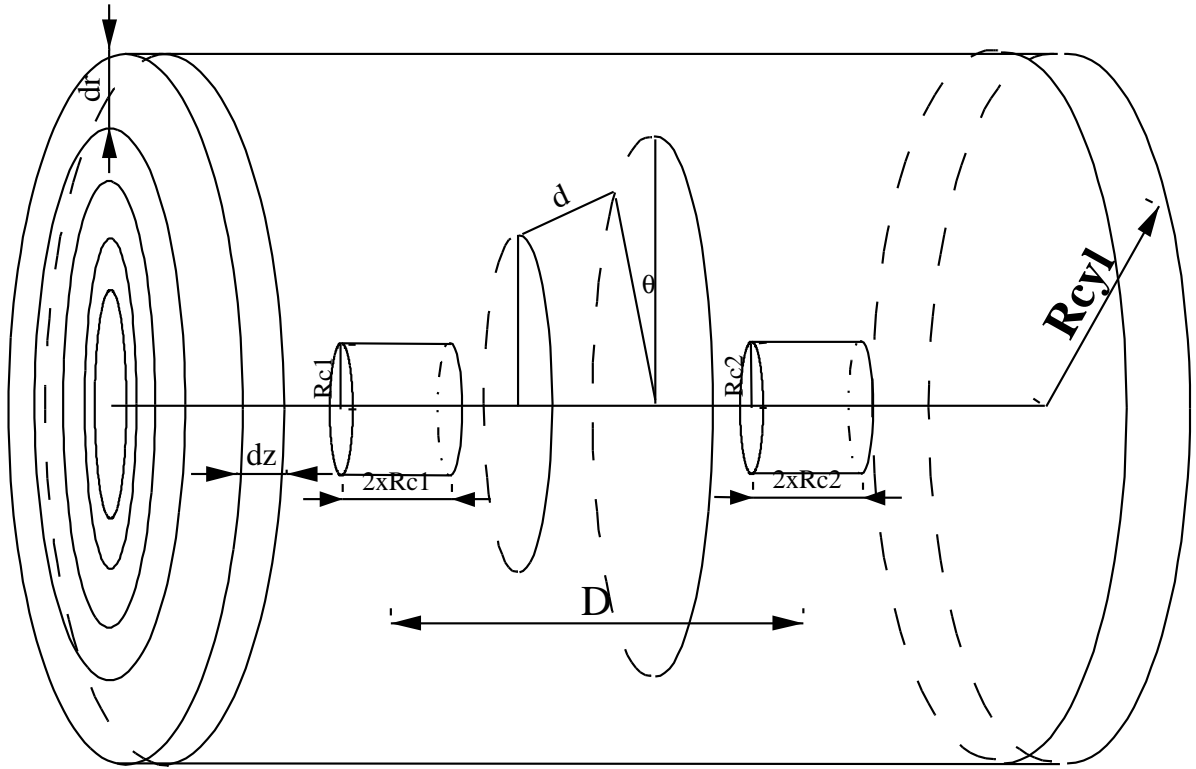}
\caption{
The geometry of the cylindrical cloud considered in the
present study. The embedded energy sources  lie
along the axis of the cylinder
(at the centers of the two small cylinders)  seperated by distance D.
The two small cylinders represent the dust free cavities near
the respective sources. 
The z--axis is defined to be along the symmetry
axis of the cylinder.
The cloud is divided into $n_{z}$ disks along z--axis, and
 each disk into $n_{r}$ concentric rings.
\label{radsch}
}
\end{figure}

\begin{figure}[p]
\plotone{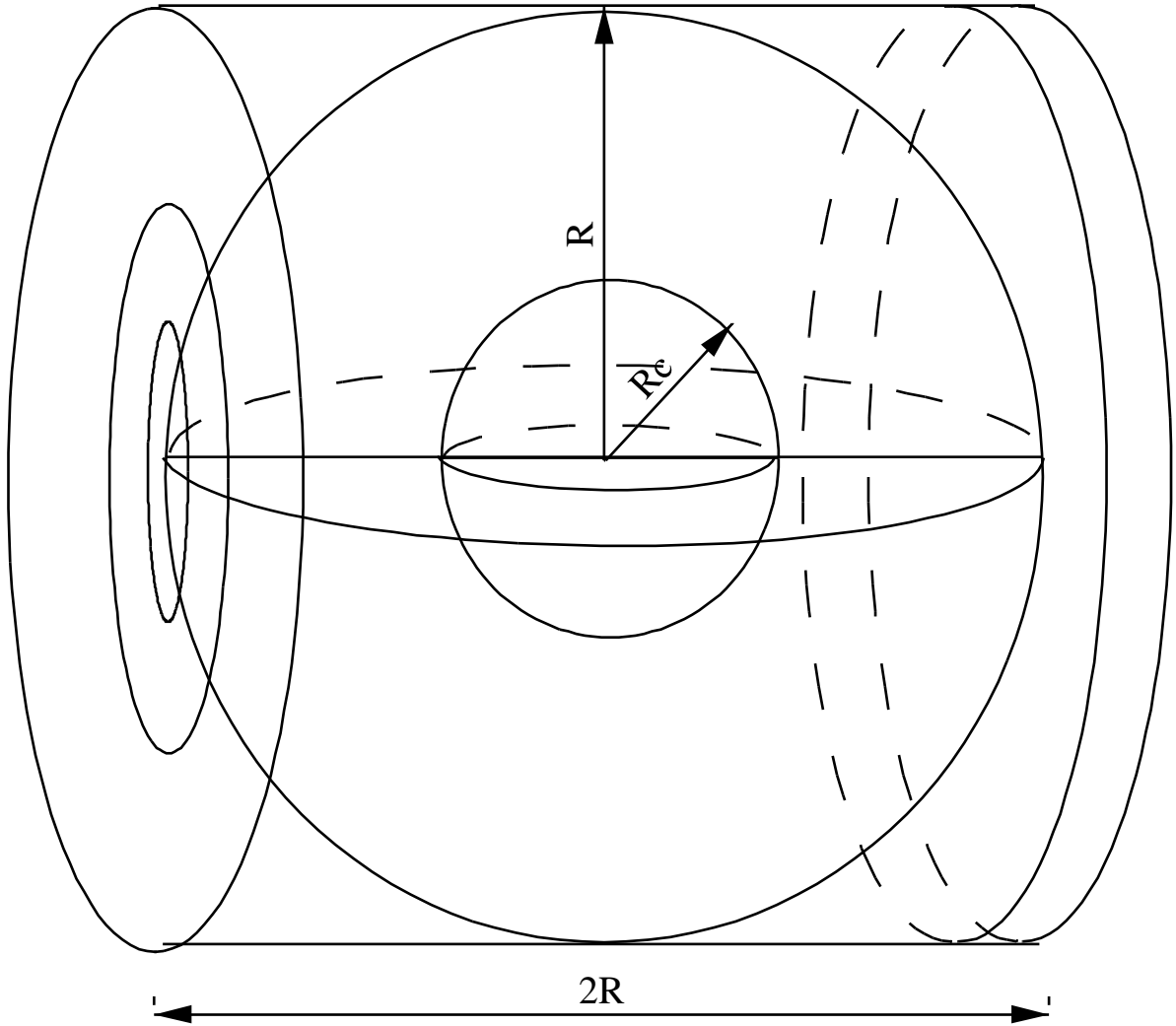}
\caption{
The geometrical dimensions of the spherical (CSDUST3) and 
the cylindrical  modelling schemes under comparison. 
\label{compmod}
}
\end{figure}

\begin{figure}[p]
\plotone{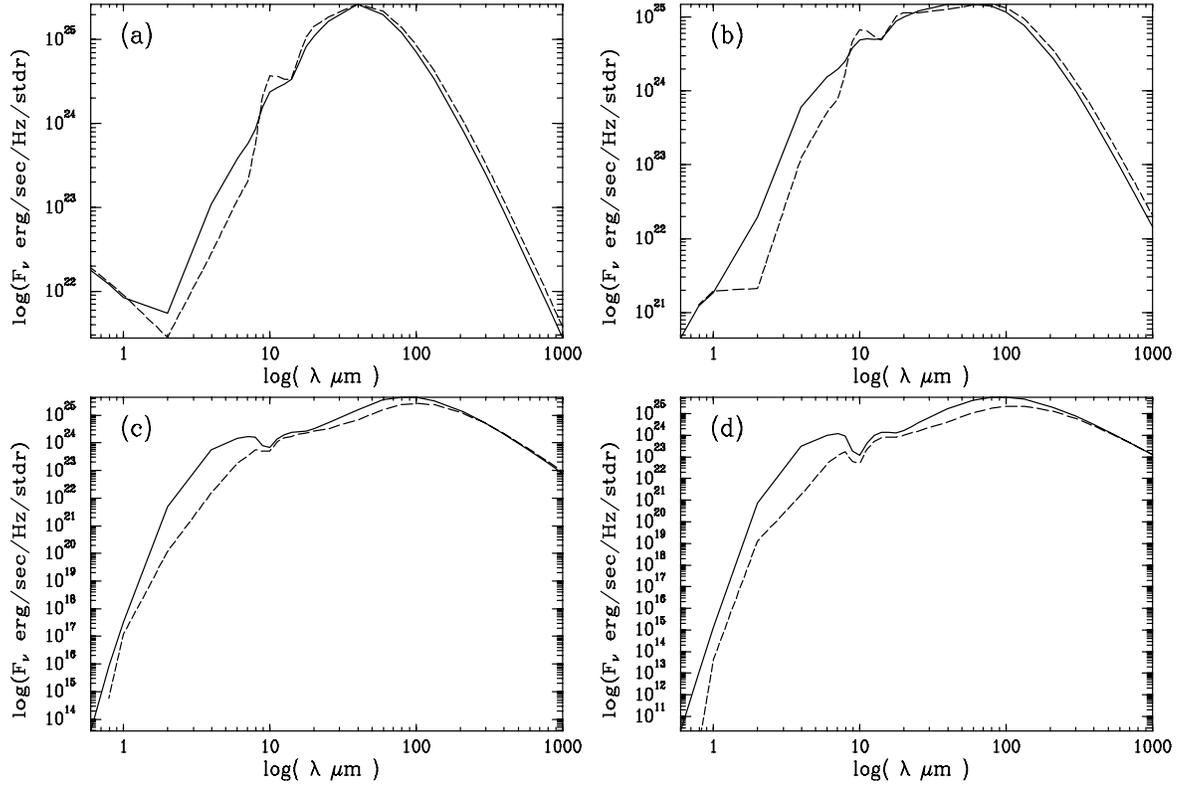}
\caption{
The comparison of predicted SEDs from the two schemes.
The continuous lines represent
CSDUST3 while the dashed lines refer to our scheme.
The SEDs correspond to $\tau_{100}$ of -- (a) 0.001, (b) .01, 
(c) .06 and (d) 0.1.
\label{seds} 
}

\end{figure}

\begin{figure}[p]
\plotone{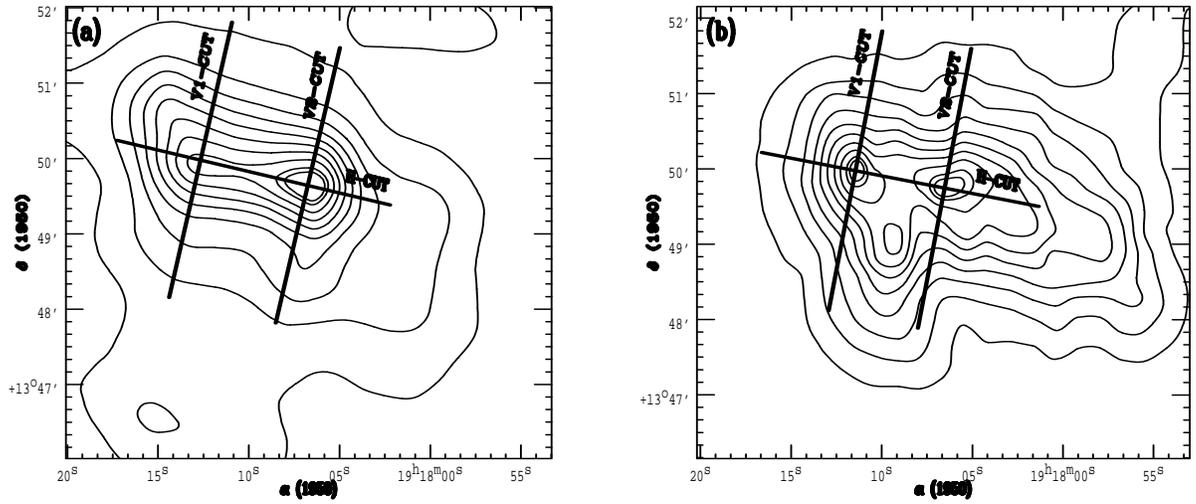}
\caption{
The infrared maps of  IRAS 19181+1349 for (a) 12 \micron\ (b) 210 \micron. 
Contours are drawn at  10, 20, 30, 40, 50, 60, 70, 80, 90, 95
percent of peak intensities 26 and 303 Jy/(arc min)$^2$ respectively.
The straight lines show the positions of radial and axial cuts
compared with the model in Fig. \ref{rc12210}. 
\label{map12210}
}

\end{figure}

\begin{figure}
\plotone{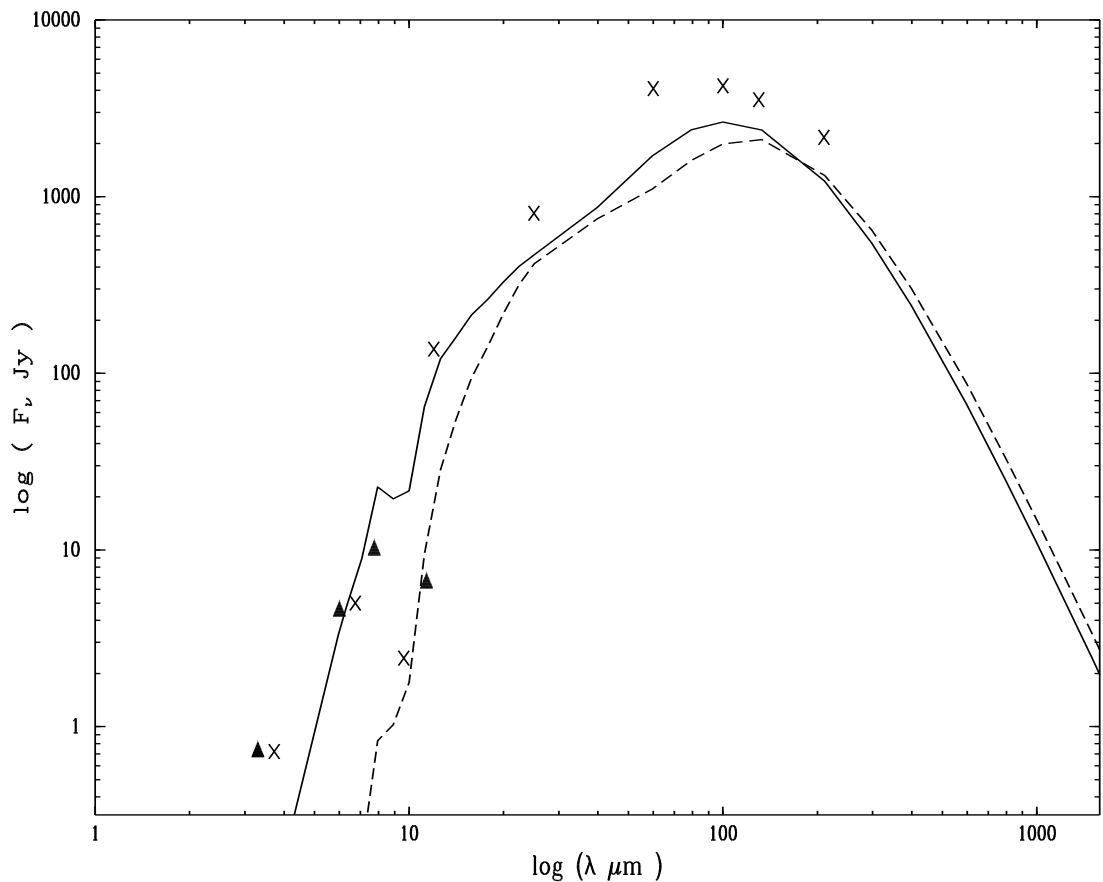}

\caption{
The SEDs for the source IRAS 19181+1349 predicted by the radiative
transfer models. 
The solid and the dashed lines represent best fit models
from the $M_{SED}$  and  $M_{RC}$ approaches, respectively.
The crosses represent HIRES (processed IRAS), TIFR and 
ISOCAM continuum flux densities. 
The solid triangles represent ISOCAM flux densities 
measured through filters centered on the PAH features.
\label{msedrc}
}

\end{figure}

\begin{figure}[p]

\plotone{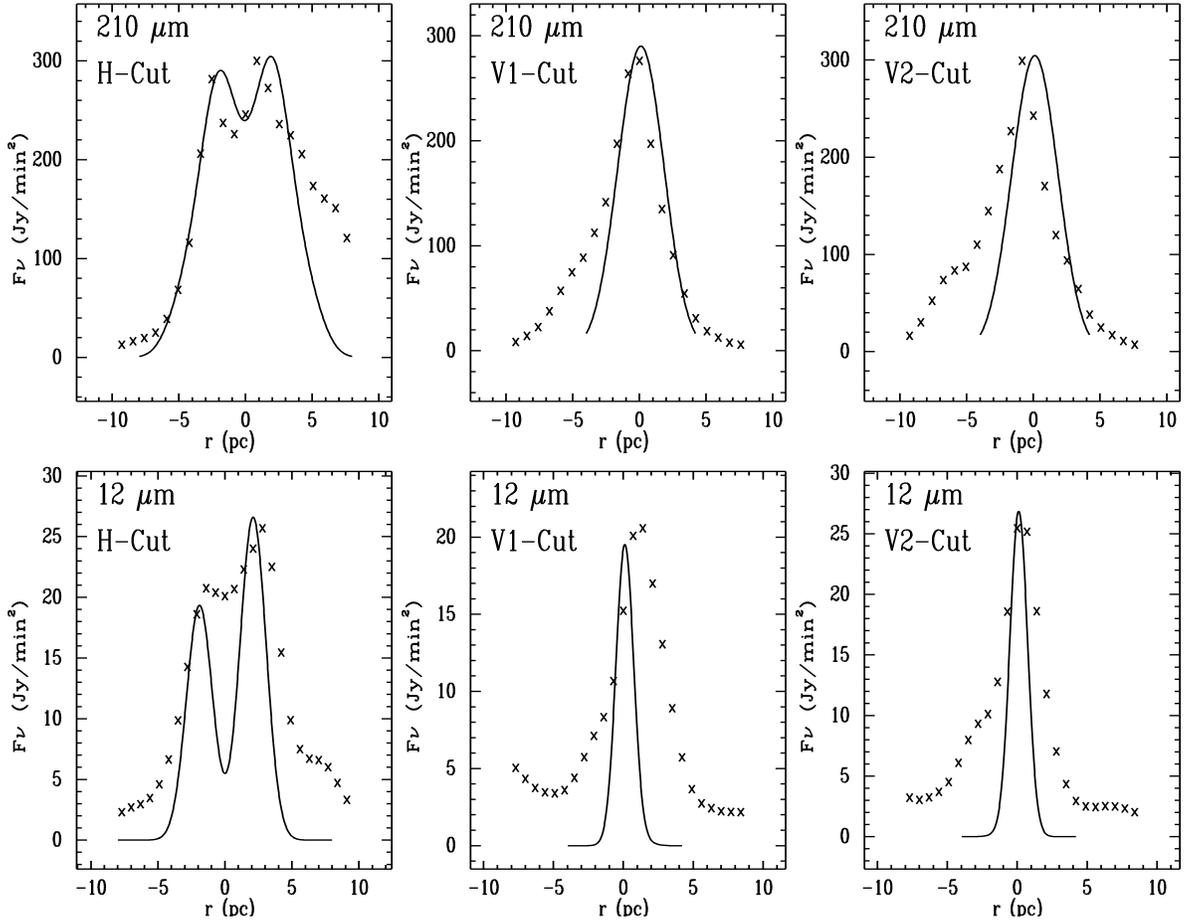}

\caption{
The axial and radial cuts predicted by the 
radiative transfer model ($M_{RC}$)
(continuous lines), compared with the observations (crosses), 
at 12   and 210 \micron.  
The spatial positions of the cuts are displayed in
Fig. \ref{map12210}.
\label{rc12210}
}
\end{figure}
\end{document}